\documentclass[12pt]{article}
\usepackage{epsfig}
\usepackage{amsmath}
\usepackage{cite}
\usepackage{amssymb}
\newlength{\defbaselineskip}
\newcommand{\setlinespacing}[1]%
           {\setlength{\baselineskip}{#1 \defbaselineskip}}

\begin{document}

\title{The effect of deformation and vibration on the alpha decay half-life}

\author{O. N. Ghodsi, E. Gholami \\
\\
{\small {\em  Department of Physics, Faculty of Science, University of Mazandaran}}\\
{\small {\em P. O. Box 47415-416, Babolsar, Iran}}\\
}
\date{}
\maketitle

\begin{abstract}
\noindent In this work, we expand upon our previous study, the
effect of  surface vibrations (low-lying vibrational states)  on the
calculation of  penetration probability in $\alpha$ decay of
spherical isotopes [Phys.Rev.C91,034611(2015)]. To this aim, the
Coulomb and proximity potential model, taking in to account the
ground state deformations of the involved nuclei along with the
surface vibrations in the daughter nucleus, is used to evaluate the
alpha decay probability. The results are compared with those
obtained by spherical potential barrier, which shows the dramatic
effect of employing the  ground state deformations  in case of
deformed nuclei. As well, including of surface vibrations give rise
to an increase in the value of tunneling probability in better
agreement with experimental data.

\end{abstract}

%======================================================================================
\newpage

{\noindent \bf{I. INTRODUCTION}}\\

Alpha decay  is an important mode of  radioactive decay to  provide
reliable information on   the nuclear structure and to the
identification of new elements[1-3]. So, experimental and
theoretical calculation of alpha decay half-life has maintained
particular
interest  in nuclear physics.\\
A simple theoretical way to determine the alpha decay width  is by
assuming the alpha particle to penetrate through a one dimensional
potential barrier; and the penetration probability  can be obtained
in terms of the Wentzel-Kramers-Brillouin (WKB) semiclassical
approximation [4-7].The knowledge  of this  potential barrier which
consists of long range coulomb interaction, short range nuclear
interaction  and centrifugal term  is essential  for the  reasonable
prediction of alpha decay half life. Different approximations give
rise to a variety of models with different accuracy. So far
different  theoretical  models such as shell model, the fission like
model, and the cluster model have been used  to determine the
potential barrier governing the alpha decay process [8-15]. The role
of some important factors like the shell effects  have been studied
by Tonozuka and  Arima [16] and the fine structure of alpha decay
has been reported by Santhosh et al.[17]. Moreover, Royer taking
account of the role of angular momentum,  proposed analytic formulas
for alpha decay half life[18]. Deformation is also an important
factor and various models incorporated the effects of deformation
and orientation  on the interaction potential [12,15,17,19]. In the
present work, in order to consider the effect of deformation and
orientation degrees of freedom, we perform calculations  in the
frame work of coulomb and proximity potential model for deformed
nuclei [20].\\
In a recent study [21], we considered  the vibrational low lying
states of nuclei with spherical equilibrium shape in their ground
states, and studied the effect of these deformations on penetration
probability by making use of two different approaches- semiclassical
and quantum mechanical. As well, It is well-known that in case of a
deformed parent  there are many  accessible ground states and low
lying  excited states in the daughter nucleus; So  in the next step
we aim to consider the coupling of these states during tunneling
process through the deformed potential barrier.\\
To this aim, at first, taking the  coulomb and proximity potential
model proposed by Shi and Swiatecki [22] as  interacting barrier,
the involved  nuclei  are considered as  spheres; and the effect of
vibrational low lying states of daughter nuclei on penetrability
through this spherical barrier  is  calculated. Then, in the
following, the ground state deformations ( $\beta_{2}$ and
$\beta_{4}$) of parent and daughter nuclei  are also incorporated to
improve the above mentioned   interaction potential. This modified
version of coulomb and proximity potential model for deformed nuclei
(CPPMDN) then is used for the calculation of alpha decay penetration
probability and to investigate the effect of low lying  excited
states. This provides an opportunity to see the effect of ground
state deformations as well as the vibrational low lying states of
daughter nuclei on the half life time calculations.

This paper is organized as follows: In section II the formalism of
interaction potentials used and the formulas of calculating alpha
decay half life are briefly described; and in the last part of this
section, the effects of vibrational states on the decay process are
considered. The results are given in
section III.\\
\\
%======================================================================================
\\
\noindent{\bf {II. THEORETICAL FORMALISM}}\\
The potential barrier governing the alpha emission is a key
component in calculation of penetration peobability.The present
paper deals with two cases: (i) all the involved nuclei treated as
spherical, (ii) the parent and daughter nuclei have an axially
symmetric deformation.  In the second case the potential barrier
depend on the the polar angle between the symmetry axis of parent or
daughter nuclei  and the direction of alpha emission. The half life
of alpha decay process is then calculated in the frame work of
quantum mechanical tunneling  using the WKB approximation.\\
The half-life time is calculated as :
\begin{equation} \label{1}
T_{\frac{1}{2}}=\frac{\ln2}{\lambda}=\frac{\ln2}{P \nu},
\end{equation}
where $ \lambda$, the decay constant, is simply the product of the
barrier penetration probability (P) and  the assault frequency
$(\nu)$. The assault frequency, the number of assaults on the
barrier per second  is evaluated from empirical zero point vibration
energy $E_{\nu}$ in the harmonic oscillator approximation. $E_{\nu}$
for alpha decay is proportional to Q value as : $ E_{\nu}= 0.095 Q $
[23]. Q is the released energy at alpha decay. The penetration
probability of alpha particle is
taken equal to one. We have used the method of evaluation of fusion probability for the calculation of tunneling probability in alpha decay.\\
P is evaluated by averaging the penetration probability over the
polar angle between the symmetry axis of axially-symmetric deformed
parent or daughter nuclei  and the direction of alpha emission,
\begin{equation} \label{2}
P=\int_{0}^{\pi/2}P_{\theta}Sin(\theta)d\theta.
\end{equation}
$P_{\theta}$, the  penetration probability in the WKB approximation
can be written as :
\begin{equation} \label{3}
P_{\theta}=\frac{1}{1+\exp[\frac{2}{\hbar}\int_{a(\theta)}^{b(\theta)}\sqrt{2\mu(V(L,\theta)-Q)}dr]}
,
\end{equation}

where $a(\theta)$ and $b(\theta)$ are the turning points obtained
from the equation $V(L,\theta)= Q$. $ \mu $ is the reduced mass,
$\mu=m\frac{A_{\alpha}A_{d}}{A_{\alpha}+A_{d}}$, where \textit{m} is
the nucleon mass; $A_{\alpha}$ and $A_{d}$ are the mass
numbers of alpha particle and daughter nucleus respectively.\\

 In Coulomb and proximity potential model for
deformed nuclei the interaction potential $V(L,\theta)$ is
considered as the sum of coulomb repulsion, nuclear proximity
potential and centrifugal parts for the post-contact
configuration.The potential barrier  for the overlap region  is then
constructed by a smooth, power-law interpolation between the contact
and parent configuration [20].

\begin{equation} \label{4}
V(L,\theta)=V_{c}(r,\theta)+V_{p}(z)+\frac{l(l+1)\hbar^{2}}{2\mu r^{2}}~~~~~~~~~ L > L_{c}
\end{equation}
\begin{equation} \label{5}
~~~V(L,\theta)=a(L-L_{0})^{n}~~~~~~~~~~~~~~~~~~~~~ L_{0}<L <L_{c}
\end{equation}

L is the overall length of the configuration, $L_{c}$ is the length
of the contact configuration  and $L_{0}$ is the diameter of the
parent nuclei. \textit{r} is the distance between fragment centers  and z is
the distance between the near surfaces of the fragments. \textit{l} represents the angular momentum and $\mu$ is the reduced mass.The
constants \textit{a} and \textit{n} are determined by the
requirement of
smooth fit of two potentials at the touching point.\\
 $V_{p}(z)$ is the proximity potential, given by

\begin{equation} \label{6}
V_{p}(z)= 4\pi b \gamma
\frac{C_{1}(\theta)C_{2}}{C_{1}(\theta)+C_{2}}\phi(\frac{z}{b}),
\end{equation}
$\gamma$ is the nuclear surface tension coefficient :
\begin{equation} \label{7}
 \gamma=0.9517~[1 - 1.7826 ~[\frac{N-Z}{A}]^{2}].
\end{equation}
N, Z and A are the neutron, proton and mass numbers of parent
nucleus. $C_{1}(\theta)$ and $C_{2}$ are the radius vectors of the
daughter
and alpha particle respectively.\\
 $\phi(\frac{z}{b}$ ) the
universal function is given as [24]:

\begin{equation} \label{8}
\phi(\xi)\approx -4.41\exp(-\xi/0.7176)~~~~~~~~~ \xi\geq1.9475
\end{equation}

\begin{equation} \label{9}
\phi(\xi)\approx-1.7817+0.9270\xi+0.0169\xi^{2}-0.05148\xi^{3}~~~~~~~~~~~~~~~0\leq\xi\leq
1.9475
\end{equation}
$\xi=\frac{z}{b}$, where b is the diffuseness of the nuclear surface
$(b\approx 1 fm)$. See
Ref.[20] for more details.\\

The Coulomb interaction between the two deformed and oriented
nuclei:
\begin{equation} \label{10}
V_{c}(r,\theta)=\frac{Z_{1}Z_{2}e^{2}}{r}+3Z_{1}Z_{2}e^{2}\sum_{\lambda,
i=1,2}\frac{1}{2\lambda+1}\frac{(R_{i}(\alpha_{i}))^{\lambda}}{r^{\lambda+1}}Y_{\lambda}^{0}(\alpha_{i})[\beta_{\lambda
i}+\frac{4}{7}\beta_{\lambda i}^{2}Y_{\lambda}^{0}(\alpha_{i})
\delta_{\lambda,2}]
\end{equation}
where,
\begin{equation} \label{11}
R_{i}(\alpha_{i})= R_{0i}[1+\sum_{\lambda}\beta_{\lambda
i}Y_{\lambda}^{0}(\alpha_{i})]
\end{equation}
and
\begin{equation} \label{12}
R_{0i}= 1.28 A_{i}^{1/3}-0.76+0.8A_{i}^{-1/3}.
\end{equation}

The ground state deformations  of the involved nuclei ( $\beta_{2}$ and $\beta_{4}$ ) are incorporated in the evaluation of the total potential.
According to the above mentioned formalism, using WKB approximation
the penetration probability has been calculated from Eq.(3).

In the following to address the effects of accessible collective
modes and the dependence of penetration probability on the low-lying
vibrational excitations, the coupled channel formalism is employed
by including all the relevant channels and assuming the harmonic
oscillator for vibrational coupling. The stationary coupled
Schrodinger equation can be written as [25]:
\begin{equation} \label{13}
[- \frac{\hbar^{2}}{2\mu} \frac{d^{2}}{dr^{2}}
+\frac{J(J+1)\hbar^{2}}{2 \mu r^{2}}+ V_N^{0}(r)+\frac{Z_{P} Z_{T}
e^{2}}{r}+\varepsilon_{n}-E]\psi_{n}(r)+\sum_{m}
V_{nm}(r)\psi_{m}(r)=0,
\end{equation}

where $V_{nm}$ are the elements of coupling Hamiltonian [25], and
$V_N^{0}$ is the potential between two interacting nuclei. Here a
Woods-Saxon potential which fitted to the coulomb and proximity
potential model in each configuration is used for the interaction
potential. The excitation energy appears as $\varepsilon_{n}$. The
above coupled channel equations are solved under the condition that
there are only incoming waves at $r=r_{min}$, the starting point of
integration, which is taken as the minimum position of the coulomb
pocket and there are only outgoing waves at infinity for all
channels except the entrance channel:

\begin{equation}\label{14}
\Psi_{n}(r)\rightarrow T_{n}\exp(-i\int_{r_{m}}^{r}k_{n}(r')dr')
 ~~~~~~r\rightarrow r_{min}
\end{equation}
\begin{equation}\label{15}
~~~~~~~~~~H^{-}_{J}(K_{n}r)\delta_{n0} +R_{n} H^{+}_{J}(K_{n}r)
~~~~r\rightarrow r_\infty
\end{equation}

practically the numerical solution is matched to a linear
combination of incoming and outgoing wave function where both the
nuclear potential and coulomb coupling are sufficiently small.
Reflection and transmission coefficients in each channel are denoted
by $R_{n}$ and $T_{n}$, respectively. $H^{-}_{J}$ and $H^{+}_{J}$
are the incoming and outgoing coulomb functions. at $r = r_{max}$,
by superposing of the incoming and outgoing coulomb waves:
\begin{equation}\label{16}
\chi_{nm}(r)=C_{nm} H^{-}_{J}(K_{m}r)+D_{nm}H^{+}_{J}(K_{m}r)
~~~~~r\rightarrow r_{max}
\end{equation}
and the solution of the coupled channel equations with the proper
boundary condition Eq.(14) and Eq.(15) are:

\begin{equation}\label{17}
\Psi_{m}(r)=\sum_{n} T_{n}\chi_{nm}(r),
\end{equation}
and at $r_{max}$,
\begin{equation}\label{18}
 \Psi_{m}(r_{max})=\sum_{n} T_{n}\chi_{nm}(r_{max}),
 \end{equation}
 by comparing with Eq.(15):
\begin{equation}\label{19}
\sum_{n}T_{n}C_{nm}=\delta_{m0},
\end{equation}
then, the penetrability correspond to every angle is given by:
\begin{equation}\label{20}
P_{\theta}=\sum_{n}\frac{K_{n}(r_{min})}{K_{0}}|T_{n}|^{2}.
\end{equation}
See Ref.[25] for more details.\\
 In this study  in order to identify the role of coupling effects on the calculation of penetration probability,
 the CCFULL code is used, considering one phonon excitations of $2^{+}$ states of daughter nuclei.
  Since the $\alpha $ particle is a closed-shell nucleus and has a high lying excited state of 20 MeV,
 its excitations are not included.
 deformation parameters and excitation energies at $2^{+}$ excitation state of daughter nuclei  used in the input file of CCFULL code are given in Table 1.\\
 \\

 %==========================================================================================================
\noindent{\bf {III. RESULTS}}\\
As pointed out in the Introduction,in a previous study the role of
coupled channel effects demonstrated in case of nuclei with
spherical equilibrium shape  in their ground state [21]. In the
present investigation the Coulomb and proximity potential  model  is
applied to the calculation of alpha decay half-life  of $^{162}Hf$,
$^{174}Os$, $^{226}Ra $, $^{226}Th$, $^{226}U$, $^{176,178,190}Pt$

and $^{188,190}Pb $ isotopes. The alpha emitters are taken such that
 the deformation and intrinsic properties such as  vibrations and
corresponding deformations to be considered.  As a first step, all
the involved nuclei are considered spherical and penetrability
through a spherical barrier is calculated based on semiclassical WKB
approximation. In the second step considering the crucial role of
ground state deformations( $\beta_{2}$ and $\beta_{4}$ ) of parent
and daughter nuclei, we extend this spherical potential barrier to
the Coulomb and proximity potential model for deformed
nuclei(CPPMDN); The potential barrier so constructed incorporates
ground state deformations, and half-life time value is found to
decrease in comparison to the spherical case (see columns two and
four of table II). In order to see the influence of vibrational
excitations on penetration probability, the coupled channel approach
is used considering the effects of coupling to the low-lying $2^{+}$
excitation state of daughter  nuclei in both the above mentioned
cases i.e spherical and deformed potential. The deformation
parameters of these nuclei used in CCFULL code taken from Ref.[26],
are given in Table I.In table II the  logarithm of half-life time
calculated from  one dimensional potential and coupled channel
approach for both the spherical and deformed potentials are
presented. It is evident  that  including of surface vibrations has
dramatic influence on the half-life value. In fact these couplings
give rise to an increase in the value of tunneling probability  due
to the reduction of potential barrier height.
Alpha decay half-lives of the above mentioned nuclei have been calculated in many recent studies. Denisov and Khudenko [27]evaluated alpha decay half-lives of 344 nuclei with the help of  empirical relations in the frame work  of unified model for alpha decay and alpha capture (UMADAC) ; Xu and Ren [28] presented a systematic calculation using microscopic density dependent cluster model(DDCM); And by using generalized liquid drop model (GLDM) Bao et al.[29] investigated the role of shell effects in the behaviour of decay half-lives. Corresponding values of half-lives from Refs.[27-29] along with the experimental values are presented in the last four columns of Table 2.From the table it may
be seen that by considering the ground state deformations (CPPMDN)
the results get modified in comparison to spherical potential (CPPM)
and  by taking into account the ground state deformations along with
surface vibrations(CPPMDN + CC)
shows better agreement with experimental data.\\

%=====================================================================================%==================================================

%========================================================================================================
\newpage
Table 1. Deformation parameters and  excitation energies at
$J^{\pi}=2^{+}$ excitation state of daughter nuclei used in the
Coupled Channel calculations [26].
\begin{center}
\begin{tabular}{c c c c c c c c c c}
  \hline
  \hline
  Nuclei & $E_{ex.}(MeV)$ & $ \beta_{vib}$ \\
   \hline
 \\
 $^{158}Yb$    & 0.3582 & 0.194  \\
 \\

   \hline
   \\
 $^{170}W$    &0.1568 & 0.24  \\
 \\

   \hline
   \\
 $^{222}Rn$    &0.1862 & 0.141  \\
 \\

   \hline
   \\
 $^{222}Ra$    &0.111 & 0.192  \\
 \\
   \hline
   \\
 $^{222}Th$    &0.183 & 0.153  \\
 \\
   \hline
   \\
 $^{172}Os$    &0.227 & 0.225  \\
 \\
   \hline
   \\
 $^{174}Os$    &0.158 & 0.226  \\
 \\
   \hline
   \\
 $^{186}Os$    &0.137 & 0.2  \\
 \\
   \hline
   \\
 $^{184}Hg$    &0.366 & 0.16  \\
 \\
   \hline
   \\
 $^{186}Hg$    &0.405 & 0.132  \\
 \\
   \hline
\end{tabular}
\end{center}

\newpage
Table 2. The decimal logarithm of alpha decay half life  through one
dimensional potential (ODP) and coupled channel approach (CC) using
two versions of coulomb and proximity potential model: spherical
(CPPM) and deformed (CPPMDN). Experimental values along with
corresponding results predicted by the unified model for alpha decay
and alpha capture (UMADAC)[27], density-dependent cluster
model(DDCM)[28], and generalized liquid drop model (GLDM)[29] are
also presented.

\begin{center}
\begin{tabular}{c c c c c c c c c c c c c c c c c c c c c c}
  \hline
  \hline
  & && \scriptsize CPPM & & & & &\scriptsize CPPMDN &  &  & \scriptsize UMADAC & \scriptsize  DDCM & \scriptsize GLDM & \scriptsize Exp. \\ \\
  \\

  & & \scriptsize ODP & &\scriptsize CC($2^{+}$) & & & \scriptsize ODP & & \scriptsize CC($2^{+}$)  & & & & & &\\
   \hline
   \hline
 \\
  \small $^{162}Hf$ & &  \footnotesize 6.878 &  & \footnotesize 6.579  & & & \footnotesize 6.599 & & \footnotesize 6.2   &  &\footnotesize 5.86 &\footnotesize --- &\footnotesize 6.17 &\footnotesize 5.8 \\
 \\
   \hline
   \\
  \small $^{174}Os$  & &  \footnotesize 6.233  & & \footnotesize 5.796  & & & \footnotesize 5.803 & & \footnotesize 5.224  &  &\footnotesize 5.27 &\footnotesize 5.38 &\footnotesize 5.447 &\footnotesize 5.34 \\
 \\
   \hline
   \\
 \small $^{226}Ra$  &  &  \footnotesize  12.393  & & \footnotesize 12.01  & & &  \footnotesize 11.697 & & \footnotesize 11.213  &  &\footnotesize 11.28 &\footnotesize --- &\footnotesize 11.164 &\footnotesize 10.73\\
 \\
 \hline
   \\
 \small $^{226}Th$  & &  \footnotesize 4.566 & & \footnotesize 4.201  &  & & \footnotesize 4.02 & &\footnotesize 3.504  &  &\footnotesize 3.58 &\footnotesize --- &\footnotesize 3.44 &\footnotesize 3.39\\
 \\
 \hline
 \\
 \small $^{226}U$  & & \footnotesize 0.395 & & \footnotesize 0.141  &  & & \footnotesize 0.022 & &\footnotesize -0.377  &  &\footnotesize -0.18 &\footnotesize --- &\footnotesize -0.49 &\footnotesize -0.57\\
 \\
 \hline
 \\
 \small $^{176}Pt$  & & \footnotesize 2.040 & & \footnotesize 1.658  &  & & \footnotesize 1.746 & &\footnotesize 1.239  &  &\footnotesize 1.26 &\footnotesize 1.415 &\footnotesize 1.415 &\footnotesize 1.22\\
 \\
 \hline
 \\
 \small $^{178}Pt$  & &  \footnotesize 3.477 & & \footnotesize 2.992  &  & & \footnotesize 3.054 & &\footnotesize 2.410  &  &\footnotesize 2.56 &\footnotesize 2.813 &\footnotesize 2.76 &\footnotesize 2.45\\
 \\
 \hline
 \\
 \small $^{190}Pt$  & & \footnotesize 20.381 & & \footnotesize 19.9747  &  & & \footnotesize 19.935 & &\footnotesize 19.4359  &  &\footnotesize 19.22 &\footnotesize --- &\footnotesize 19.255 &\footnotesize 19.31\\
 \\
 \hline
 \\
 \small $^{188}Pb$  & &  \footnotesize 2.829 & & \footnotesize 2.585  &  & & \footnotesize 2.710 & &\footnotesize 2.407  & &\footnotesize 2.22 &\footnotesize --- &\footnotesize 2.04 &\footnotesize 2.06\\
 \\
 \hline
 \\
 \small $^{190}Pb$  & &  \footnotesize 4.7581 & & \footnotesize 4.535  &  & & \footnotesize 4.637 & &\footnotesize 4.401  &  &\footnotesize 4.13 &\footnotesize --- &\footnotesize 3.88 &\footnotesize 4.25\\
 \\
 \hline
\end{tabular}
\end{center}

\end{document}